\documentstyle[aps,prl,psfig,epsf,floats]{revtex} 

\draft

\begin{document}
\twocolumn[\hsize\textwidth\columnwidth\hsize\csname
@twocolumnfalse\endcsname 

\title{Physics of Synchronized Neutrino Oscillations
Caused by Self-Interactions}

\author{Sergio Pastor$^{(1)}$, Georg Raffelt$^{(1)}$, Dmitry V.~
Semikoz$^{(1,2)}$}

\address{$^{(1)}$Max-Planck-Institut f\"ur Physik
(Werner-Heisenberg-Institut), F\"ohringer Ring 6, 80805 M\"unchen,
Germany}

\address{$^{(2)}$Institute for Nuclear Research of the Academy of
Sciences of Russia,\\ 60th October Anniversary Prospect 7a, Moscow
117312, Russia}

\date{February 5, 2002}

\maketitle
                    
\begin{abstract}
In the early universe or in some regions of supernovae, the neutrino
refractive index is dominated by the neutrinos themselves.  Several
previous studies have found numerically that these self-interactions
have the effect of coupling different neutrino modes in such a way as
to synchronize the flavor oscillations which otherwise would depend on
the energy of a given mode. We provide a simple explanation for this
baffling phenomenon in analogy to a system of magnetic dipoles which
are coupled by their self-interactions to form one large magnetic
dipole which then precesses coherently in a weak external magnetic
field. In this picture the synchronized neutrino oscillations are
perfectly analogous to the weak-field Zeeman effect in atoms.

\end{abstract}

\vskip2.0pc]

%%%%%%%%%%%%%%%%%%%%%%%%%%%%%%%%%%%%%%%%%%%%%%%%%%%%%%%%%%%%%%%%%%%%%%
%% Section I %%%%%%%%%%%%%%%%%%%%%%%%%%%%%%%%%%%%%%%%%%%%%%%%%%%%%%%%%
%%%%%%%%%%%%%%%%%%%%%%%%%%%%%%%%%%%%%%%%%%%%%%%%%%%%%%%%%%%%%%%%%%%%%%

\section{Introduction}

\begin{figure}
\centerline{\psfig{file=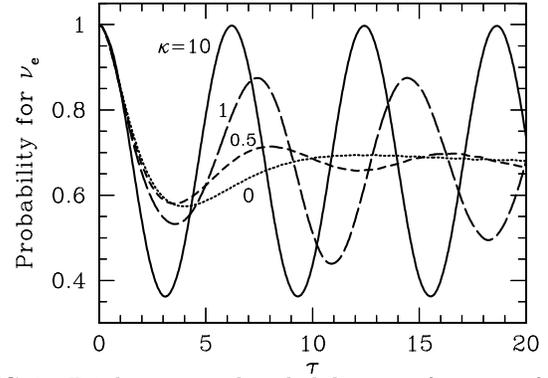,angle=0,width=7cm}}
\caption{Total $\nu_e$ survival probability as a function of time,
where $\tau \equiv (\Delta m^2/2p_0)t$ and $p_0 = \langle p^{-1}
\rangle^{-1} \simeq 2.2\,T$. The curves are for different values
$\kappa$ of the neutrino self-coupling as indicated where $\kappa=0$
corresponds to vacuum oscillations.}
\label{totalP}
\end{figure}

In a two-flavor system of mixed neutrinos, the flavor content of a
given state oscillates with the frequency $\Delta m^2/2p$ where
$\Delta m^2 = m_2^2-m_1^2$ is the neutrino mass-squared difference and
$p$ is the momentum. Therefore, if a neutrino ensemble encompasses
many modes with many different momenta, these modes develop growing
relative phases so that the overall flavor content of the ensemble
quickly decoheres. This trivial effect is illustrated in
Fig.~\ref{totalP} for an ensemble of neutrinos (no anti-neutrinos)
with a thermal momentum distribution at temperature $T$. The vacuum
mixing angle was taken to be $\sin2 \theta = 0.8$ and all neutrinos
were originally in a pure $\nu_e$ state. In our example the momentum
distribution is very broad so that the flavor decoherence takes place
within about one oscillation period (dotted line in
Fig.~\ref{totalP}).

This behavior changes dramatically when the neutrinos feel a
significant weak-interaction potential caused by the presence of the
other neutrinos. We express the strength of the neutrino-neutrino
potential in terms of the parameter
\begin{equation}
\kappa \equiv
\frac{2\sqrt{2}G_{\rm F}n_\nu p_0}{\Delta m^2}
\end{equation}
where $p_0\equiv\langle p^{-1}\rangle^{-1}$.  When the neutrino-neutrino
potential is comparable or much larger than a typical $\Delta m^2/2p$,
corresponding to $\kappa={\cal O}(1)$ or larger, the modes get locked
to each other---the entire ensemble oscillates with a common frequency
$\omega_{\rm synch}$ which corresponds to a certain average of $\Delta
m^2/2p$ (Fig.~\ref{totalP}).  This stunning effect was first
discovered in numerical studies of early-universe neutrino
oscillations~\cite{Samuel:1993uw} and then elaborated and applied in a
large series of
papers~\cite{Kostelecky:1994ys,Kostelecky:1993yt,Kostelecky:1993dm,%
Kostelecky:1995dt,Kostelecky:1995xc,Samuel:1996ri,Kostelecky:1996bs,%
Pantaleone:1998xi}.

We note that the mode synchronization effect discussed
in~\cite{Bell:2001kq} is unrelated to our present case.  When frequent
flavor-blind collision occur on a time scale much faster than the
oscillation period, then the energy of a given neutrino is averaged
over an oscillation period, leading to a common oscillation frequency.

Our present mode synchronization effect illustrated in
Fig.~\ref{totalP} is a strictly nonlinear effect caused by the
neutrino-neutrino self-interactions and as such seems difficult to
understand. We presently develop a very simple and physically
transparent theory of this effect, taking full advantage of the
equivalence of our problem with the spin-precession of a magnetic
dipole in magnetic fields.  It will become clear that the essential
effect of the neutrino self-coupling is to lock the individual
neutrino modes to form one large ``magnetic moment'' which then
spin-precesses in a weak external field. Therefore, the equation of
motion returns to a simple linear form.

Our approach provides a transparent and intuitive analytic framework
which nicely accommodates and illuminates the results of the previous
literature which were largely based on numerical studies.  Once the
framework is established, it is easy to study various generalizations
and special cases that otherwise would be difficult to predict or
understand.

In Sec.~II we begin by setting up the equations of motion for neutrino
oscillations with the inclusion of a neutrino-neutrino potential. In
Sec.~III we develop the picture of coupled magnetic moments to explain
the synchronization effect. In Sec.~IV we include anti-neutrinos, a
situation where the system can behave qualitatively different from the
neutrino-only case. Finally we summarize our findings in Sec.~V.

\section{Equations of Motion}

Our starting point is the well-known spin-precession picture for
neutrino oscillations in 
vacuum~\cite{Smirnov:ij,Bouchez:1986kb,Ermilova:ph,Stodolsky:1987dx}
\begin{equation}
\dot{\bf P}=\frac{\Delta m^2}{2p}\,{\bf B}\times{\bf P}.
\label{vacuum}
\end{equation}
Here, ${\bf P}$ is the polarization vector in flavor space of a
neutrino mode with momentum $p$.  In the usual way, the $z$-component
of ${\bf P}$ gives us the probability for finding the neutrino, say,
in the electron flavor state by virtue of ${\rm
prob}(\nu_e)=\frac{1}{2}(1+P_z)$. The vector ${\bf
B}=(\sin2\theta,0,-\cos2\theta)$ with the mixing angle $\theta$ gives
us an effective ``magnetic field'' around which ${\bf P}$ precesses.
Therefore, ${\bf P}$ plays the role of an angular momentum vector
while ${\bf M}=(\Delta m^2/2p)\,{\bf P}$ plays the role of a magnetic
dipole moment associated with ${\bf P}$.  The quantity $\Delta
m^2/2p=|{\bf M}|/|{\bf P}|$, gives us the proportionality between
${\bf M}$ and ${\bf P}$ and thus plays the role of the ``gyromagnetic
ratio'' for a given mode, determining the rate of
precession.

In much of the literature, the equation of motion is written in the
form $\dot{\bf P}={\bf V}\times{\bf P}$ without distinguishing clearly
between the effective angular momentum ${\bf P}$ and its associated
magnetic moment. For our present discussion this distinction is
crucial. Still, we could split ${\bf V}$ in different ways between the
gyromagnetic ratio, the unit of magnetic moment $\mu$, and the ${\bf
B}$-field which really stands for $\mu {\bf B}$. For example, we might
have used $p_0/p$ as the gyromagnetic ratio with $p_0$ some typical or
average momentum, and defined ${\bf B}=\mu{\bf B}=(\Delta
m^2/2p_0)\,(\sin2\theta,0,-\cos2\theta)$.  However, we have preferred
to avoid introducing an additional quantity $p_0$, and it is
convenient to define ${\bf B}$ as a unit vector.

We will frequently consider the polarization vector for an entire
ensemble of neutrinos, 
\begin{equation}
{\bf J}\equiv\sum_{j=1}^{N_\nu} {\bf P}_j,
\end{equation}
i.e.\ we consider a large volume ${\cal V}$ filled homogeneously with
$N_\nu$ neutrinos. We also assume an isotropic distribution of the
momenta so that it suffices to specify the modulus of the momentum of
a given mode, $p_j=|{\bf p}_j|$.

There is no closed equation of motion for ${\bf J}$ because the
individual modes oscillate with different frequencies. Evidently,
however, the projection of ${\bf J}$ on ${\bf B}$ is conserved. On the
other hand, the fast precession of the individual mode polarizations
around ${\bf J}$ average the transverse components of the individual
modes to zero so that the asymptotic value ${\bf J}_\infty=({\bf
B}\cdot {\bf J}){\bf B}$ obtains. (Recall that in our definition ${\bf
B}$ is a unit vector.)  For maximum mixing, and if initially all
neutrinos were in a flavor eigenstate, ${\bf J}_{\infty}=0$,
corresponding to an incoherent mixture of both flavors.

A background medium consisting, say, of protons, neutrons and
electrons modifies the ``magnetic field.'' Assuming our two-flavor
system involves electron neutrinos the substitution is
\begin{equation}
\frac{\Delta m^2}{2p}\, {\bf B}\to \frac{\Delta m^2}{2p}\, {\bf B}
+ \sqrt{2} G_{\rm F}\, n_e\,{\bf \hat z}
\end{equation}
with $n_e$ the electron number density. Therefore, the precession is
no longer around a common direction for all modes. If we started with
a situation of maximum mixing, then the medium reduces the effective
mixing angle for all modes. For a very dense medium, the effective
magnetic field will be almost perfectly along the 
\hbox{$z$-direction}, suppressing flavor oscillations entirely.

This is very different if we consider a neutrino ensemble so dense
that the neutrinos themselves produce a significant refractive index.
In that case the medium's contribution to the refractive index is not
along the flavor direction (i.e.\ along the $z$-axis), but rather
along the direction of ${\bf J}$. Put another way, neutrinos produce
an ``off-diagonal refractive index''~\cite{Pantaleone:1992eq} because
a given background neutrino may be a coherent superposition of flavor
states. The equation of motion for a single mode $j$ now
reads~\cite{Sigl:1993fn}
\begin{equation}
\dot{\bf P}_j=\frac{\Delta m^2}{2p_j}\,{\bf B}\times{\bf P}_j
+\frac{\sqrt2 G_{\rm F}}{{\cal V}}\,{\bf J}\times{\bf P}_j,
\label{eq:motion1}
\end{equation}
where the second term represents the self-interactions.  If we sum
this equation over all modes, then the second term becomes
proportional to ${\bf J}\times{\bf J}=0$ so that
\begin{equation}
\dot{\bf J}=
{\bf B}\times\sum_{j=1}^{N_\nu}\frac{\Delta m^2}{2p_j}{\bf P}_j.
\label{eq:motion1a}
\end{equation}
Therefore, the first derivative $\dot {\bf J}$ of the ensemble's
polarization vector is not affected by the self-interactions. Still,
the evolution of ${\bf J}$ is changed because the evolution of the
individual modes is affected. However, the vacuum oscillations are not
obviously suppressed even in a dense gas of neutrinos, in contrast to
a standard background medium.

Our case of active-active neutrino oscillations is very different from
the active-sterile case ~\cite{Sigl:1993fn,McKellar:1994ja}. Sterile
neutrinos do not produce a weak potential so that there is no
off-diagonal refractive index.  The neutrino contribution to the
``effective magnetic field'' is along the $z$-direction, i.e.\ not
proportional to ${\bf J}$. Therefore, the self-interaction term is
somewhat more similar to the effect of an external background medium,
although the oscillation equations, of course, remain non-linear.

\section{Explanation of Synchronized Oscillations}

It is now easy to demonstrate that Eq.~(\ref{eq:motion1}) implies a
synchronized precession of all modes around ${\bf B}$ if the neutrino
density is sufficiently large. To this end we first imagine the
vacuum oscillation term to be absent, i.e.\ we consider the
equation of motion
\begin{equation}
\dot{\bf P}_j=
\frac{\sqrt2 G_{\rm F}}{{\cal V}}\,{\bf J}\times{\bf P}_j.
\label{eq:motion2}
\end{equation}
Every individual mode precesses around the direction of ${\bf J}$. Of
course, if all neutrinos were initially prepared in a specific flavor
state, then all ${\bf P}_j$ as well as ${\bf J}$ are aligned along the
$z$-direction, and no precession takes place (perfectly coherent
state).  Likewise, if the individual ${\bf P}_j$ point in random
directions so that ${\bf J}=0$, again there are no precessions
(perfectly incoherent flavor mixture).  We consider the general case
where the ${\bf P}_j$ initially point in many different directions,
but do not add to zero.

Next we switch on the vacuum term from Eq.~(\ref{vacuum}), i.e.\ a
weak external ``magnetic field''. With ``weak'' we mean that for a
typical mode the precession frequency around ${\bf J}$ is much larger
than the one around ${\bf B}$. This implies that the evolution of a
given mode remains dominated by ${\bf J}$. The fast precession around
${\bf J}$ implies that the projection of ${\bf P}_j$ on ${\bf J}$ is
conserved while the transverse component averages to zero on a fast
time scale relative to the slow precession around ${\bf B}$. If ${\bf
J}$ moves slowly, then the individual modes will follow ${\bf J}$. Put
another way, the individual modes are coupled to each other by their
strong ``internal magnetic fields,'' forming a compound system with
one large magnetic moment. It is this compound object which precesses
around ${\bf B}$.  Of course, if the external field is much larger
than the internal ones (dilute neutrino gas), then the modes will
decouple and precess individually around the external field with their
separate vacuum oscillation frequencies.

Our picture is perfectly analogous to the Zeeman effect in atoms. An
atomic state is characterized by its spin angular momentum ${\bf S}$,
its orbital angular momentum ${\bf L}$, and the total angular momentum
${\bf J}={\bf L}+{\bf S}$.  In a weak external magnetic field the
spin-orbit coupling caused by the internal magnetic fields
(Russell-Saunders coupling) remains intact, the external field is only
a perturbation. In this case it is the total angular momentum ${\bf
J}$ which precesses, i.e.\ which determines the atomic level
splittings caused by the external $B$-field.  On the other hand, if
the external field is much stronger than the internal one, then ${\bf
L}$ and ${\bf S}$ decouple and precess independently around the
external field: the atomic levels are determined by the separate
quantization of ${\bf L}$ and ${\bf S}$ along the external ${B}$-field
(Paschen-Back effect).

Granting that in the neutrino case the polarization vectors of the
individual modes are locked in the sense that ${\bf J}$ indeed forms
one large ``angular momentum,'' the evolution of the compound system
is governed by the equation
\begin{equation}
\dot{\bf J}=\omega_{\rm synch}\,{\bf B}\times{\bf J}.
\label{Jdot}
\end{equation}
It remains to determine the value of $\omega_{\rm synch}$ which plays
the role of the gyromagnetic ratio for the compound system.  Of the
individual modes, the external field ``sees'' only the projection
along ${\bf J}$ because the transverse components average to
zero. Therefore, the contribution of mode $j$ to the total magnetic
moment ${\bf M}$ is $(\Delta m^2/2p_j)\,{\bf\hat J}\cdot{\bf P}_j$ so
that
\begin{equation}
{\bf M}={\bf\hat J}\,\sum_{j=1}^{N_\nu}
\frac{\Delta m^2}{2p_j}\,{\bf\hat J}\cdot{\bf P}_j,
\end{equation}
where ${\bf\hat J}$ is a unit vector in the direction of ${\bf J}$.
Therefore, the gyromagnetic ratio is
\begin{equation}
\omega_{\rm synch}=\frac{|{\bf M}|}{|{\bf J}|}=
\frac{1}{|{\bf J}|}\sum_{j=1}^{N_\nu}
\frac{\Delta m^2}{2p_j}\,{\bf\hat J}\cdot{\bf P}_j.
\label{wsynchnu}
\end{equation}
In particular, if all modes started aligned (coherent flavor state)
then $|{\bf J}|=N_\nu$ and ${\bf\hat J}\cdot{\bf P}_j=1$ so that
\begin{equation}
\omega_{\rm synch}=\left\langle\frac{\Delta m^2}{2p}\right\rangle
=\frac{1}{N_\nu}\sum_{j=1}^{N_\nu}\frac{\Delta m^2}{2p_j}
\end{equation}
in agreement with~\cite{Kostelecky:1995dt}.

If an external medium is present, the uncoupled modes precess around
different ${\bf B}$-field vectors rather than a common direction. When
the modes are coupled by self-interactions, it is still the one ${\bf
J}$ which precesses around one common ${\bf B}$ field which is a
suitable average of the individual ${\bf B}_j$ which easily can be
worked out.

Returning to the simpler case of a common ${\bf B}$ for all modes, the
calculation of $\omega_{\rm synch}$ amounts to determining the
Land\'e-factor for ${\bf J}$ in the atomic analogy. This is the
problem of calculating the magnetic moment of a system if the angular
momentum is the sum of individual components which have magnetic
moments with different gyromagnetic ratios.  In this case the vector
sums of the angular momenta and of the magnetic moments are not
co-linear.  In atomic physics, the spin angular momentum produces
twice the magnetic moment of the orbital angular momentum, hence the
complication.

We stress that, contrary to the previous literature, our analysis
shows that there is nothing special about the initially aligned state,
even though this state may be most motivated by the neutrino
applications when all of them start in one flavor state.  Any initial
configuration of ${\bf J}$ precesses as one vector. Again, in atomic
physics ${\bf L}$ and ${\bf S}$ can be combined in different ways to
form one ${\bf J}$. For example, a $p$-state with $L=1$ and $S=1/2$
can combine to a $J=3/2$ or a $J=1/2$ state; there is nothing special
about the ``aligned'' state ($J=3/2$).

\begin{figure}[t]
\centerline{\psfig{file=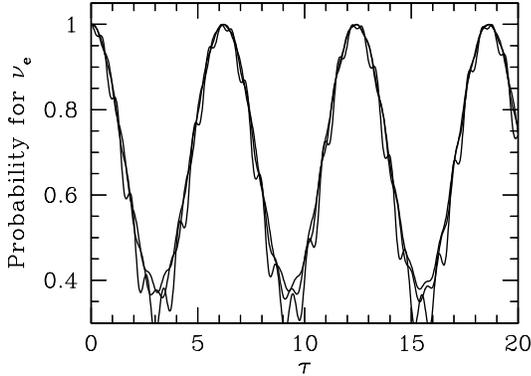,angle=0,width=7cm}}
\caption{Evolution of the $\nu_e$ survival probability for three
values of the neutrino momenta in the presence of a strong neutrino
self-potential term ($\kappa = 10$).}
\label{modes}
\end{figure}

Moreover, even though the initial ${\bf J}$ precesses as a compound
system, the individual modes do not oscillate in unison, except for
the special case of perfect initial alignment and infinitely strong
self-coupling.  In the general case the motion of every ${\bf P}_j$ is
a fast precession around ${\bf J}$, superimposed on a slow precession
of ${\bf J}$ around ${\bf B}$. The compound motion of ${\bf P}_j$ is
generally rather complicated and different for every mode.  This is
illustrated in Fig.~\ref{modes}, where the synchronized oscillations
of three different modes are shown for $\kappa = 10$, starting with a
perfectly aligned state.

Our treatment predicts that any set of initial ${\bf P}_j$ leads to
synchronized oscillations in the sense of a precession of the initial
${\bf J}$ with a single frequency $\omega_{\rm synch}$.  The length of
the initial-state ${\bf J}$ is conserved while the oscillation
frequency depends on details of the initial distribution of
polarization vectors.  The system does not prefer one particular
synchronized state over another. For example, the completely
incoherent state will stay that way, and will not spontaneously align
the ${\bf P}_j$ to form a coherent state, in agreement with the
stability analysis of~\cite{Pantaleone:1998xi}.  Any initially
prepared ${\bf J}$ will do nothing but precess about ${\bf B}$.  Put
another way, the nonlinear aspect of the neutrino system manifests
itself in the coupling of the individual ${\bf P}_j$ to each other to
form a compound ${\bf J}$ which acts as one large angular momentum.
Beyond this, the system is easily understood in terms of a linear
equation of motion.

Our analysis is also entirely independent of the number of neutrinos
or modes. For example, if there are only two modes nothing in our
analysis changes so that the synchronization effect should not be
viewed as a collective phenomenon.  In fact, the atomic example was
one consisting of two coupled magnetic moments, the orbital and spin
terms.  We have checked with a simple numerical code that a system of
two or three modes indeed behaves as expected according to our
treatment. Actually, a two-mode system can be fully solved
analytically so that in principle the physical essence of the
synchronization effect can be calculated without recourse to numerical
methods.

There are exceptions to our general statements which will be of
interest in the next section. Evidently we can construct
``pathological'' cases where several angular momenta ${\bf P}_j$ add
up to a vanishing or arbitrarily small ${\bf J}$ while the magnetic
moments do not, leading to a gyromagnetic ratio which can be
constructed to become arbitrarily large. In this case $\omega_{\rm
synch}$ will no longer represent some typical $\Delta m^2/2p$, but can
be constructed to become arbitrarily large. In this case the motion of
${\bf J}$ is not necessarily slow compared to the internal precessions
of the individual ${\bf P}_j$ so that our treatment is no longer
adequate.  However, the perfectly incoherent state of an ensemble of
many modes will not be pathological in this sense because the random
distribution of ${\bf P}_j$ will ensure that both $|{\bf J}| \sim
|{\bf M}| \sim 1/\sqrt{N_\nu}$, where $N_\nu$ is the number of
modes. In the limit $N_\nu \rightarrow \infty$ both ${\bf J}$ and
${\bf M}$ vanish simultaneously.

\section{Neutrinos Plus Antineutrinos}

As a next step we may extend our analysis to the case where neutrinos
and antineutrinos are simultaneously present, an inevitable situation
in a realistic system such as the early universe unless the neutrino
chemical potentials are extremely large. To first order in $G_{\rm F}$
the equations of motion are~\cite{Sigl:1993fn}
\begin{eqnarray}
\dot{\bf P}_j&=&+\frac{\Delta m^2}{2p_j}\,{\bf B}\times{\bf P}_j
+\frac{\sqrt2 G_{\rm F}}{{\cal V}}\,({\bf J}-{\bf\overline J})
\times{\bf P}_j
\nonumber\\
\dot{\bf\overline P}_k&=&
-\frac{\Delta m^2}{2p_k}\,{\bf B}\times{\bf\overline P}_k
+\frac{\sqrt2 G_{\rm F}}{{\cal V}}\,({\bf J}-{\bf\overline J})
\times{\bf\overline P}_k,
\label{nu-anti-nu}
\end{eqnarray}
where overbarred quantities refer to anti-neutrinos. In particular,
the total polarization vectors are
\begin{equation}
{\bf J}=\sum_{j=1}^{N_\nu} {\bf P}_j\qquad\hbox{and}\qquad
{\bf\overline J}=\sum_{k=1}^{N_{\bar\nu}} 
{\bf\overline P}_k.
\end{equation}
As explained in~\cite{Sigl:1993fn},
the definition of the polarization vector for anti-neutrinos
is ``reversed'' in the sense that in vacuum it precesses in the
opposite direction of neutrinos.\footnote{In the flavor oscillation
case one can avoid this minus sign by defining both polarization
vectors in the same way so that the vacuum oscillation equations look
the same.  In this case the sign change has to be introduced by hand
in the non-linear potential terms by flipping the sign of one
component of the potential. This can be achieved by defining ${\bf
P}^*$ as the vector with the second component reversed, leading to
equations of motion of the form
\begin{eqnarray}
\dot{\bf P}_j&=&+\frac{\Delta m^2}{2p_j}\,{\bf B}\times{\bf P}_j
+\frac{\sqrt2 G_{\rm F}}{{\cal V}}\,({\bf J}-{\bf\overline J}^*)
\times{\bf P}_j,
\nonumber\\
\dot{\bf\overline P}_k&=&
+\frac{\Delta m^2}{2p_k}\,{\bf B}\times{\bf\overline P}_k
-\frac{\sqrt2 G_{\rm F}}{{\cal V}}\,({\bf J}^*-{\bf\overline J})
\times{\bf\overline P}_k.\nonumber
\end{eqnarray}
This approach was used in much of the literature.  While it is
mathematically equivalent to our treatment, we think that it obscures
the simplicity of the equations because their vector form is
destroyed.}  This corresponds to fermions with a true magnetic moment
which have opposite gyromagnetic ratios for particles and
anti-particles.  For example, a neutron and an anti-neutron
spin-precess in opposite directions in the same external magnetic
field.

It is obvious that a system consisting of neutrinos and anti-neutrinos
behaves the same way as one consisting of neutrinos only, except that
the role of the total angular momentum is now played by ${\bf I}={\bf
J}-{\bf\overline J}$.  The anti-particles appear as normal modes of
the system, except that they sport negative gyromagnetic ratios.  It
is ${\bf I}={\bf J}-{\bf\overline J}$ which precesses slowly around
the external field, while all ${\bf P}_j$ and ${\bf\overline P}_k$
remain pinned to ${\bf I}$. The corresponding evolution equation for 
the compound system is
\begin{equation}
\dot{\bf I}=\omega_{\rm synch}\,{\bf B}\times{\bf I}.
\end{equation}
where the total gyromagnetic ratio is
\begin{equation}
\omega_{\rm synch}=
\frac{1}{|{\bf I}|}\left (\sum_{j=1}^{N_\nu}
\frac{\Delta m^2}{2p_j}\,{\bf\hat I}\cdot{\bf P}_j+
\sum_{k=1}^{N_{\bar\nu}}
\frac{\Delta m^2}{2p_k}\,{\bf\hat I}\cdot{\bf \overline P}_k\right )
\label{wsynchantinu}
\end{equation}
We have checked with a numerical code several situations with a
thermal population of neutrinos and antineutrinos, and the results
were always as expected.

Therefore, a system consisting of neutrinos and anti-neutrinos
typically behaves qualitatively similar to the neutrino-only
case. However, the negative gyromagnetic ratios of anti-neutrinos
relative to neutrinos allow for ``pathological'' situations where the
system behaves qualitatively differently than above.

One such case is a thermal ensemble without chemical potential, i.e.\
a situation where $N_{\nu}=N_{\bar\nu}$.  If all neutrinos start in a
given flavor state we have ${\bf I}={\bf J}-{\bf\overline J}=0$. This
situation is ``pathological'' in the sense that a vanishing or very
small ${\bf I}$ is associated with a large magnetic moment because
particles and anti-particles enter with exactly opposite magnetic
moments. To illustrate this case we consider only one mode of
particles and anti-particles so that the equations of motion are
\begin{eqnarray}
\dot{\bf P}&=&+\omega\,{\bf B}\times{\bf P}
+({\bf P}-{\bf\overline P})\times{\bf P},
\nonumber\\
\dot{\bf\overline P}&=&
-\omega\,{\bf B}\times{\bf\overline P}
+({\bf P}-{\bf\overline P})\times{\bf\overline P},
\label{nu-antinu-example}
\end{eqnarray}
with a suitable $\omega$. This implies
\begin{equation}
\dot{\bf P}-\dot{\bf\overline P}
=\omega{\bf B}\times ({\bf P}+{\bf\overline P}).
\end{equation}
\begin{figure}[t]
\centerline{\psfig{file=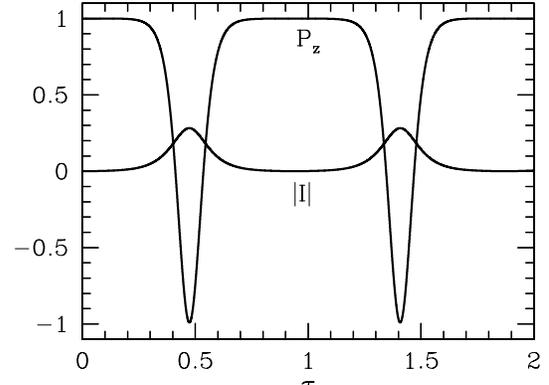,angle=0,width=7cm}}
\caption{Evolution of $P_z$ and $|{\bf I}|$ for
the system of Eqs.~(\ref{nu-antinu-example}), when $\sin 2
\theta=0.01$ and in the presence of a strong neutrino self-potential
with $\omega = -0.01$, corresponding to an inverted mass hierarchy
$\Delta m^2 <0$.}
\label{patholog}
\end{figure}

Evidently ${\bf I}={\bf P}-{\bf\overline P}$ is not conserved if the
two polarization vectors start aligned so that at first ${\bf I}=0$.
The effect of the external field is to drive ${\bf P}$ and
${\bf\overline P}$ apart, creating a net ${\bf I}$ which is orthogonal
to ${\bf B}$. In the case of large mixing both ${\bf P}$ and
${\bf\overline P}$ will oscillate much faster than they would in
vacuum, yet convert to the other flavor. 

In the case of small mixing the result strongly depends on the sign of
$\omega$ in Eqs.~(\ref{nu-antinu-example}), i.e.~on the sign of
$\Delta m^2$. For $\omega > 0$, the directions of
the vectors ${\bf P}$, ${\bf \overline P}$ and ${\bf B}$ almost
coincide.  The role of the strong neutrino self-potential term is just
to increase the oscillation frequency, while the amplitude of $P_z$ is
the same as in the vacuum case. 

For an inverted mass hierarchy ($\Delta m^2<0$) and small mixing
angle, ${\bf B}$ is close to the $z$-axis, but it is almost opposite
to the initial directions of ${\bf P}$ and ${\bf \overline P}$.  A
small seed of $|{\bf I}| \neq 0$ is enough to drive ${\bf P}$ and
${\bf \overline P}$ to the opposite direction from their initial
orientation, i.e.~one can achieve complete flavor conversion.  We
illustrate this situation in Fig.~\ref{patholog}, where the
$z$-component of the polarization vector and the modulus of ${\bf I}$
are plotted for $\omega = -0.01$ and $\sin 2 \theta = 0.01$. One can
see that $P_z$ evolves from $1$ to $-1$ and back. This resonance has a
different origin from the Mikheev-Smirnov-Wolfenstein mechanism
because it comes from the neutrino self-potential term.

These phenomena were discovered and discussed
in~\cite{Kostelecky:1993dm} and subsequent papers of that series.  We
merely stress that our way of writing the equations allows for a
straightforward visualization of the motion of the polarization
vectors.

Another physically motivated ``pathological'' case is when two flavors
of neutrinos are thermally populated with large but opposite chemical
potentials. In the early universe this corresponds to a hypothetical
initial condition of a vanishing or small lepton-number density, yet
an anomalously large density of radiation in the form of neutrinos.
Since the chemical potentials are assumed to be opposite we have ${\bf
J}=-{\bf \overline J}$ so that ${\bf I}={\bf J}-{\bf \overline
J}=2{\bf J}$ is now ``large.'' On the other hand $\omega_{\rm
synch}=0$ so that flavor oscillations take place with a vanishing
frequency, i.e.\ the initial condition is frozen without further
evolution. This appears to be the only case where the large
neutrino-neutrino self-potential acts to prevent flavor oscillations.

These ``pathological'' cases have an atomic counterpart in the form of
positronium where the two spins are associated with equal but opposite
magnetic moments. Therefore, the $I=1$ state (ortho-positronium) has a
vanishing magnetic moment and thus shows no Zeeman splittings in a
weak external field. Likewise, the $I=0$ state (para-positronium)
consists of only one level and thus cannot split.  Therefore, even
though we have a system of two spins associated with two magnetic
moments, there is no weak-field Zeeman effect.  In a strong field the
electron and positron spins and magnetic moments are separately
quantized along the external $B$-field, giving rise to a nontrivial
level structure.

\section{Conclusions}

In summary, we have provided a simple and physical explanation of the
synchronized oscillations observed in the numerical treatment of dense
neutrino ensembles. The effect is perfectly analogous to the coupling
of several angular momenta, for example spin and orbital angular
momentum in an atom, to form one large compound angular momentum with
one large associated magnetic dipole moment which precesses as one
object in a weak external field. Anti-neutrinos are naturally included
in our picture if one observes that they carry ``negative gyromagnetic
ratios'' in flavor space. The nonlinear nature of neutrinos
oscillating in a background of neutrinos is thus reduced to a very
simple and well-known coupling effect of magnetic moments to each
other.  Our picture provides a transparent framework that accounts
perfectly for all of the previously discussed synchronization
phenomena in the literature, and that allows one to derive both
general properties and special cases in a unified framework.

The practical impact of synchronized neutrino oscillations in the
early universe will be studied in a forthcoming paper~\cite{BBN}.

%%%%%%%%%%%%%%%%%%%%%%%%%%%%%%%%%%%%%%%%%%%%%%%%%%%%%%%%%%%%%%%%%%%%%%
%% Acknowledgments %%%%%%%%%%%%%%%%%%%%%%%%%%%%%%%%%%%%%%%%%%%%%%%%%%%
%%%%%%%%%%%%%%%%%%%%%%%%%%%%%%%%%%%%%%%%%%%%%%%%%%%%%%%%%%%%%%%%%%%%%%

\section*{Acknowledgments}

We thank G.~Sigl for helpful comments.  In Munich, this work was
partly supported by the Deut\-sche For\-schungs\-ge\-mein\-schaft
under grant No.\ SFB 375 and the ESF network Neutrino Astrophysics. SP
was supported by a Marie Curie fellowship of the European Commission
under contract HPMFCT-2000-00445.

%%%%%%%%%%%%%%%%%%%%%%%%%%%%%%%%%%%%%%%%%%%%%%%%%%%%%%%%%%%%%%%%%%%%%%
%% References %%%%%%%%%%%%%%%%%%%%%%%%%%%%%%%%%%%%%%%%%%%%%%%%%%%%%%%%
%%%%%%%%%%%%%%%%%%%%%%%%%%%%%%%%%%%%%%%%%%%%%%%%%%%%%%%%%%%%%%%%%%%%%%

\end{document}